\documentclass[12pt,preprint]{aastex}
\usepackage{xcolor}

\begin{document}

\title{A Solar-type Stellar Companion to a Deep Contact Binary in a Quadruple System}

\author{X. Zhou\altaffilmark{1,2,3}, S.-B. Qian\altaffilmark{1,2,3}, J. Zhang\altaffilmark{1,2}, L.-Q. Jiang\altaffilmark{4}, B. Zhang\altaffilmark{1,2,3}, J. Kreiner\altaffilmark{5}}

\singlespace

\altaffiltext{1}{Yunnan Observatories, Chinese Academy of Sciences (CAS), P. O. Box 110, 650216 Kunming, China; zhouxiaophy@ynao.ac.cn}
\altaffiltext{2}{Key Laboratory of the Structure and Evolution of Celestial Objects, Chinese Academy of Sciences, P. O. Box 110, 650216 Kunming, China}
\altaffiltext{3}{University of Chinese Academy of Sciences, Yuquan Road 19\#, Sijingshang Block, 100049 Beijing, China}
\altaffiltext{4}{Department of Physics, School of Science, Sichuan University of Science \& Engineering, Zigong, 643000, China}
\altaffiltext{5}{Mt. Suhora Astronomical Observatory, Pedagogical University of Cracow, Poland}
\begin{abstract}
The four-color ($B$ $V$ $R_c$ $I_c$) light curves of V776 Cas are presented and analyzed using the Wilson-Devinney (W-D) method. It is discovered that V776 Cas is an early F-type (F2V) overcontact binary with a very high contact degree ($ f=64.6\,\%$) and an extremely low mass ratio ($q=0.130$), which indicate that it is at the final evolutionary stage of cool short-period binaries. The mass of the primary and secondary stars are calculated to be $M_1 = 1.55(\pm0.04)M_\odot$, $M_2 = 0.20(\pm0.01)M_\odot$. V776 Cas is supposed to be formed from an initially detached binary system via the loss of angular momentum due to the magnetic wind. The initial mass of the present primary and secondary components are calculated to be $M_{1i} = 0.86(\pm0.10)M_\odot$ and $M_{2i} = 2.13(\pm0.04)M_\odot$. The observed-calculated ($O$-$C$) curve exhibits a cyclic period variation, which is due to the light-travel time effect (LTTE) caused by the presence of a third component with a period of 23.7 years. The mass of the third component is estimated to be $M_3 = 1.04(\pm0.03)M_\odot$ and the orbital inclination of the third component is calculated to be $i' = 33.1^{\circ}$. The distance of the binary system to the mass center of the triple system is calculated to be $a'_{12} = 3.45AU$. The presence of the close-in tertiary component may play an important role in the formation and evolution of this binary system by drawing angular momentum from the central system.

\end{abstract}

\keywords{binaries : close --
          binaries : eclipsing --
          stars: evolution --
          stars: individual (V776 Cas)}

\section{Introduction}
V776 Cas (BD+69 0121, HIP 8821, $V$ = $9^{m}.09$) is the brighter member of the visual binary ADS 1485. The companion, at a separation of $5''.38$, is 2 mag fainter than the contact binary. V776 Cas was discovered as an eclipsing binary with a small amplitude by the $Hipparcos$ mission \citep{1997ESASP.402...13L}. \citet{1997IBVS.4513....1D} gave its spectral type to be F0, color index to be $B - V = 0.525$ mag and variability ranging from 8.943 to 9.090 mag in $V$ band, but it gave the period of the binary system to be P = 0.8808 days, which was nearly twice of the actual period (P = 0.44041574 days). The binary star was observed photometrically by \citet{1999IBVS.4702....1G} who assumed the star to be an EW type undergoing marginal eclipses or an ellipsoidal variable. Their $V$ band light curve observed by using a 14-cm telescope in Piera Observatory showed that the primary minimum was 0.019 mag deeper than the secondary one. \citet{2001AJ....122.1974R} obtained the first spectroscopic elements of the system: mass ratio $q = 0.130\pm0.004$, $V_\gamma = -24.71\pm0.69$ km s$^{-1}$, $(M_1+M_2)sin^3i = 0.975\pm0.026M_\odot$. They pointed out that V776 Cas was an A-subtype W UMa binary system with a spectral type of F2V. After that, photometric analysis of V776 Cas had been carried out by \citet{2004NewA....9..425D} (\citet{2004ASPC..318..192E}), \citet{2005AcA....55..389Z} and \citet{2007ASPC..362...82O}. \citet{2006AJ....132..650D} discovered that V776 Cas was a triple system with a solar-type tertiary component, and the spectroscopic observations gave the fitted temperature to be $T_3 = 6100K$. They also gave the mass ratio to be $M_3/M_1 = 0.67$.  \citet{2012JKAS...45....1G} also analyzed the radial velocity curves of V776 Cas and obtained its orbital parameters, which were almost consistent with those obtained by \citet{2001AJ....122.1974R}.

There are many different types of close binaries and there isn't a single formation scenario for all types. Two evolutionary mechanisms are proposed for this problem, which are nuclear evolution and angular momentum evolution \citep{1988MNRAS.231..341H}. \citet{1976ApJS...32..583W}'s model suggested that the primary component in the binary system expanded first and filled the Roche lobe on its way to the red giant branch, then the mass transfer occurred and led the initially detached binary system to be a contact one via a case A mass transfer. However, angular momentum evolution via magnetic braking \citep{1962AnAp...25...18S,1968MNRAS.138..359M} was thought as the main mechanism for the formation and evolution of W UMa-type contact binaries by \citet{1970PASJ...22..317O}, \citet{1979A&A....80..287V} and \citet{1982A&A...109...17V}. The possible scenario is that W UMa-type binaries are formed from low-mass detached close binaries through the loss of angular momentum due to the magnetic wind, and the binary systems will finally merged, thus blue straggles are formed \citep{2004ARep...48..219T}. A new method was established to compute the initial mass and orbital evolution of W UMa-type contact binaries \citep{2013MNRAS.430.2029Y,2014MNRAS.437..185Y}. The results suggested that binary systems with initial mass (the initially more massive one) higher than $1.8M_\odot$ evolved to be A-subtype systems while systems with initial mass lower than this became W-subtype systems. For the binary systems with initial mass higher than $1.8M_\odot$, the nuclear evolution is the principal mechanism responsible for the Roche lobe filling process, while the binary systems with initial mass less than $1.8M_\odot$ evolve into the contact phase mainly due to the rapid angular momentum evolution. More detailed analyses of W UMa-type binary systems are needed to test the theory in the future.

V776 Cas is a typical A-subtype W UMa-type overcontact binary system with a close-in stellar component and a distant visual component. Orbital properties of the close-in tertiary component will provide valuable information on the formation of close binaries and stellar dynamical interaction. However, the orbital period variations of V776 Cas have been neglected since it was discovered, which is a very important part for the research of contact binary stars. As discussed by \citet{2010MNRAS.405.1930L}, the most plausible explanation of the cyclic period changes in close binaries is the light-travel time effect (LTTE) caused by the presence of the tertiary component around the binary system. And the survey of \citet{2006A&A...450..681T} implied that the close-in tertiary around the close binary may play an important role in the formation and evolution of close binary by shortening the orbital period of the binary system through transferring angular momentum from the central binary system. In the present paper, four color ($B$ $V$ $R_c$ $I_c$) light curves of V776 Cas are obtained and analyzed. All available times of light minimum are collected and the observed times of light minimum-calculated times of light minimum ($O - C$) curve is presented for the first time. Based on the spectroscopic elements, photometric analyses and the $O - C$ fitting results, the absolute physical parameters of the close binary system and the close-in tertiary component are obtained. It will give us a comprehensive understanding to this close binary system and be an excellent target to test the formation scenario of W UMa-type contact binaries.

\section{Observations}
The $B$ $V$ $R_c$ $I_c$ bands light curves of V776 Cas were carried out in three nights on January 10, January 16 and  February 27, 2013 with an Andor DW436 1K CCD camera attached to the 85cm reflecting telescope at Xinglong Observation Base. The coordinates of the variable star, the comparison star and the check star were listed in Table \ref{Coordinates1}. The integration time were 15s for $B$ band, 10s for $V$ band, 6s for $R_c$ band, and 5s for $I_c$ band, respectively. The light curves of those observations were displayed in Fig. 1.

During the observations, the broadband Johnson-Cousins $B$ $V$ $R_c$ $I_c$ filters were used. PHOT (measured magnitudes for a list of stars) of the aperture photometry package in the IRAF \footnote {The Image Reduction and Analysis Facility is hosted by the National Optical Astronomy Observatories in Tucson, Arizona at URL iraf.noao.edu.} was used to reduce the observed images.

\begin{table}[!h]
\begin{center}
\caption{Coordinates of V776 Cas, the comparison, and the check stars.}\label{Coordinates1}
\begin{small}
\begin{tabular}{ccccc}\hline\hline
Targets          &   name               & $\alpha_{2000}$           &  $\delta_{2000}$         &  $V_{mag}$     \\ \hline
Variable         &   V776 Cas           &$01^{h}53^{m}23^{s}.4$     & $+70^\circ02'33''.4$     &  $9.09$           \\
The comparison   &   GSC 04314-00961    &$01^{h}52^{m}32^{s}.2$     & $+70^\circ04'01''.7$     &  $11.12$     \\
The check        &   GSC 04314-00473	&$01^{h}52^{m}57^{s}.7$     & $+70^\circ04'29''.5$     &  $12.49$     \\
\hline\hline
\end{tabular}
\end{small}
\end{center}
\end{table}

\begin{figure}[!h]
\begin{center}
\includegraphics[width=13cm]{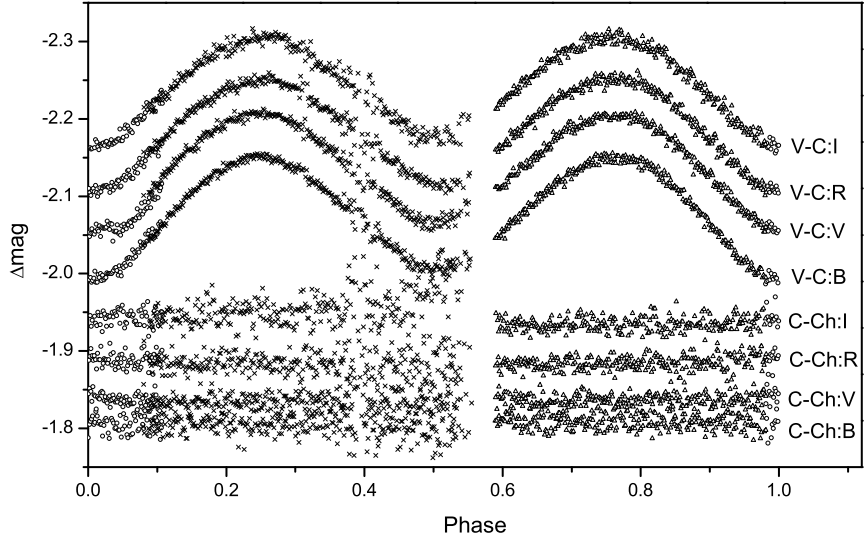}
\caption{CCD photometric light curves in $B$ $V$ $R_c$ and $I_c$ bands. The magnitude difference between the comparison and the check stars are presented. The standard deviations of the comparison-check observations are 0.012 mag for $B$ band, 0.010 mag for $V$ band, 0.012 mag for $R_c$ band and 0.019 mag for $I_c$ band. Crosses, triangles and open circles correspond to the data observed on January 10, January 16 and February 27, respectively.}
\end{center}
\end{figure}

Times of light minimum of V776 Cas were also observed and determined, which were listed in Table \ref{Newminimum}.

\begin{table}[!h]
\begin{center}
\caption{New CCD times of light minimum for V776 Cas.}\label{Newminimum}
\begin{tabular}{cccccc}\hline
    JD (Hel.)     &  Error (days)  & Min. &   Filter      & Method  &Telescopes\\\hline
  2454781.3140    & $\pm0.0003$    &   I  &   $V$         &  CCD    &    85cm   \\
  2454816.9898    & $\pm0.0003$    &   I  &   $V$         &  CCD    &    50cm   \\
  2456243.0517    & $\pm0.0042$    &   I  &   $N$         &  CCD    &    1m   \\
  2456249.2172    & $\pm0.0004$    &   I  &   $N$         &  CCD    &    60cm   \\
  2456253.1826    & $\pm0.0021$    &   I  &   $VR_cI_c$   &  CCD    &    60cm   \\
  2456350.9589    & $\pm0.0066$    &   I  &   $BVR_cI_c$  &  CCD    &    85cm   \\
\hline
\end{tabular}
\end{center}
\textbf
{\footnotesize Notes.} \footnotesize 60cm and 1m correspond to the 60cm and 1m reflecting telescope in Yunnan Observatories. 50cm and 85cm correspond to the 50cm and 85cm reflecting telescope in Xinglong Observation base.
\end{table}

\section{Orbital Period Investigation of V776 Cas}

The study of orbital period change is a very important part for contact binary system. However, the period change investigation of V776 Cas has been neglected since it was discovered. In the present work, all available times of light minimum are collected. Minimum times with the same epoch have been averaged, and only the mean values are listed in Table \ref{Minimum1}. Using the following linear ephemeris,
\begin{equation}
Min.I(HJD) = 2456249.2172+0^{d}.44041574\times{E},\label{linear ephemeris}
\end{equation}
the $O - C$ values are calculated and listed in the fourth column of Table \ref{Minimum1} and plotted in the upper panel of Fig 2. Based on the least-square method, the new ephemeris is
\begin{equation}
\begin{array}{lll}
Min. I = 2456249.2107(\pm0.0001)+0.440416842(\pm0.000000009)\times{E}
         \\+0.0109(\pm0.0001)\sin[0.^{\circ}01829\times{E}+130.^{\circ}9(\pm0.^{\circ}6)]
\end{array}
\end{equation}
The sinusoidal term reveals a cyclic change with a period of 23.7 years and an amplitude of 0.0109 days. The residuals from Equation (2) are displayed in the bottom panel of Fig. 2.

\begin{table}[!h]
\caption{$(O-C)$ values of light minima for V776 Cas.}\label{Minimum1}
\begin{center}
\small
\begin{tabular}{cclllcc}\hline\hline
JD (Hel.)      &  Min &   Epoch     & $(O-C)$      &   Error       & Method       &  Reference       \\
(2400000+)     &      &             &              &               &              &\\\hline
47941.2021     & I    &  -18864	    & 	-0.0126    &               &    CCD       &  1    \\
48025.0820     & II   &  -18673.5   &   -0.0319    &               &    CCD       &  1    \\
48072.4360     & I    &  -18566	    & 	-0.0226    &               &    CCD       &  1    \\
48160.7468     & II   &  -18365.5   &   -0.0151    &               &    CCD       &  1    \\
48226.1370     & I    &  -18217	    & 	-0.0267    &               &    CCD       &  1    \\
48370.8188     & II   &  -17888.5   &   -0.0214    &    0.0011     &    CCD       &  1    \\
48371.0347     & I    &  -17888	    &   -0.0257    &    0.0011     &    CCD       &  1    \\
48433.3500     & II   &  -17746.5   &   -0.0293    &               &    CCD       &  1    \\
48500.0850     & I    &  -17595	    &   -0.0173    &    0.0001     &    CCD       &  2    \\
48611.5045     & I    &  -17342	    &   -0.0229    &               &    CCD       &  1    \\
50814.8957     & I    &  -12339	    &   -0.0317    &               &    CCD       &  3    \\
51788.2184     & I    &  -10129	    &   -0.0278    &    0.0008     &    CCD       &  3    \\
52556.3107     & I    &  -8385	    &   -0.0205    &    0.0005     &    CCD       &  4    \\
52556.5306     & II   &  -8384.5	&   -0.0208    &    0.0004     &    CCD       &  4    \\
52932.4331     & I    &  -7531  	&   -0.0132    &               &    CCD       &  1    \\
52937.4908     & II   &  -7519.5	&   -0.0202    &               &    CCD       &  1    \\
53086.1374     & I    &  -7182  	&   -0.0140    &    0.0001     &    CCD       &  5    \\
53351.7116     & I    &  -6579	    &   -0.0105    &    0.0001     &    CCD       &  6    \\
53966.5385     & I    &  -5183  	&   -0.0040    &    0.0007     &    CCD       &  7    \\
54039.4274     & II   &  -5017.5	& 	-0.0038    &    0.0004     &    CCD       &  8    \\
54093.6015     & II   &  -4894.5	& 	-0.0009    &    0.0003     &    CCD       &  9    \\
54700.4962     & II   &  -3516.5	& 	-0.0009    &    0.0004     &    CCD       &  10    \\
54781.3140     & I    &  -3333	    &    0.0025    &    0.0003     &    CCD       &  22    \\
54788.3554     & I    &  -3317	    &   -0.0028    &    0.0002     &    CCD       &  11    \\
54816.9898     & I    &  -3252  	& 	 0.0046    &    0.0003     &    CCD       &  22    \\
54819.6290     & I    &  -3246	    &    0.0013    &    0.0002     &    CCD       &  12    \\
55154.3451     & I    &  -2486	    & 	 0.0014    &    0.0028     &    CCD       &  13    \\
55202.5705     & II   &  -2376.5	& 	 0.0013    &    0.0005     &    CCD       &  14    \\
55223.2688     & II   &  -2329.5	& 	 0.0001    &    0.0003     &    CCD       &  15    \\
55397.4509     & I    &  -1934	    &   -0.0023    &    0.0008     &    CCD       &  15    \\
55421.4527     & II   &  -1879.5	&   -0.0031    &    0.0006     &    CCD       &  15    \\
55475.4078     & I    &  -1757	    &    0.0011    &    0.0006     &    CCD       &  15    \\
55779.5150     & II   &  -1066.5	& 	 0.0012    &    0.0006     &    CCD       &  16    \\
55804.3990     & I    &  -1010	    &    0.0017    &    0.0163     &    CCD       &  17    \\
55872.6655     & I    &  -855	    &    0.0038    &    0.0004     &    CCD       &  18    \\
56155.4089     & I    &  -213	    &    0.0003    &    0.0003     &    CCD       &  19   \\
56156.5120     & II   &  -210.5	    & 	 0.0023    &    0.0008     &    CCD       &  16    \\
56220.3713     & II   &  -65.5	    & 	 0.0013    &    0.0139     &    CCD       &  20    \\
56221.2511     & II   &  -63.5	    &    0.0003    &    0.0007     &    CCD       &  16    \\
56221.4725     & I    &  -63	    &    0.0015    &    0.0003     &    CCD       &  16    \\
56243.0517     & I    &  -14	    &    0.0003    &    0.0040     &    CCD       &  22    \\
\hline
\end{tabular}
\end{center}
\end{table}

\addtocounter{table}{-1}
\begin{table}[!h]
\small
\begin{center}
\caption{Continuted}
\begin{tabular}{cclllcc}\hline\hline
JD (Hel.)      &  Min &   Epoch     & $(O-C)$      &   Error       & Method       &  Reference       \\
(2400000+)     &      &             &              &               &              &\\\hline
56249.2172     & I    &  0	        &    0         &    0.0004     &    CCD       &  22    \\
56253.1826     & I    &  9      	&    0.0017    &    0.0021     &    CCD       &  22    \\
56350.9589     & I    &  231	    &    0.0057    &    0.0070     &    CCD       &  22    \\
56500.4830     & II   &  570.5	    &    0.0086    &    0.0004     &    CCD       &  19    \\
56542.5356     & I    &  666	    &    0.0015    &    0.0003     &    CCD       &  19    \\
56638.5450     & I    &  884	    &    0.0003    &    0.0002     &    CCD       &  21    \\\hline
\end{tabular}
\end{center}
\textbf
{\footnotesize Reference:} \footnotesize (1) private provision; (2) \citet{1999IBVS.4702....1G}; (3) \citet{2001AJ....122.1974R}; (4) \citet{2003IBVS.5407....1T}; (5) \citet{2005IBVS.5592....1K}; (6) \citet{2005IBVS.5602....1N}; (7) \citet{2007IBVS.5777....1P}; (8) \citet{2006IBVS.5736....1C}; (9) \citet{2007IBVS.5760....1N}; (10) \citet{2009IBVS.5898....1P}; (11) \citet{2009IBVS.5887....1Y}; (12) \citet{2009IBVS.5875....1N}; (13) \citet{2010IBVS.5941....1H}; (14) \citet{2011IBVS.5966....1N}; (15) \citet{2011IBVS.5980....1P}; (16) \citet{2013IBVS.6044....1P}; (17) \citet{2012IBVS.6026....1H}; (18) \citet{2013OEJV..160....1H}; (19) \citet{2014IBVS.6114....1Z}; (20) \citet{2013IBVS.6070....1H}; (21) \citet{2014IBVS.6125....1B}; (22) present work;
\end{table}

\begin{figure}[!h]
\begin{center}
\includegraphics[width=13cm]{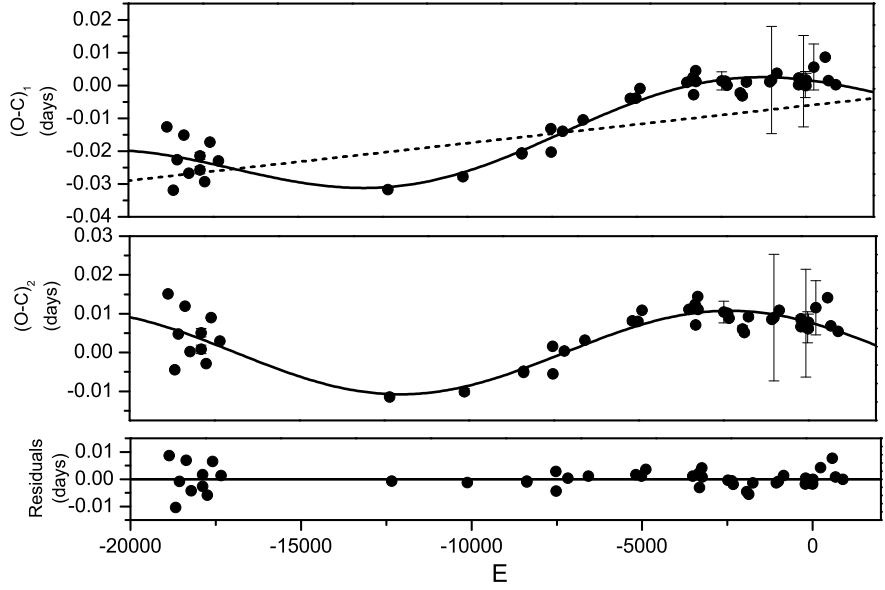}
\caption{The $(O-C)_1$ values of V776 Cas from the linear ephemeris of Equation (1) are presented in the upper panel. The solid line in the panel refers to the combination of a new linear ephemeris and a cyclic variation. The dash line represents the new linear ephemeris. In the middle panel of Fig. 2, $(O-C)_2$ values calculated from the new linear ephemeris in Equation (2) are displayed. The solid line refers to a theoretical light-travel time effect (LTTE) orbit of the tertiary component in the system. The residuals from the whole effect are displayed in the bottom panel.}
\end{center}
\end{figure}

\section{Photometric Solutions of V776 Cas}
As shown in Fig. 1, the variations of light curves in four colors are continuous and have very small magnitude differences between the depth of the primary minimum and secondary one, which indicate that V776 Cas is a typical EW-type contact binary. In Fig. 1, the light curves have been shifted vertically which will make no difference to the results of W-D program as differential photometry method is used. The phases are calculated with the following linear ephemeris:
\begin{equation}
Min.I(HJD) = 2456249.2172+0^{d}.44041574\times{E}.\label{linear ephemeris}
\end{equation}
To understand its geometrical structure and evolutionary state, the $B$ $V$ $R_c$ and $I_c$ bands light curves shown in Fig. 1 are analyzed using the W-D program of 2013 version \citep{Wilson1971,Wilson1979,Wilson1990,Van2007,Wilson2008,Wilson2010,Wilson2012}. The numbers of observed data points used in W-D program are 624 in $B$ band, 616 in $V$ band, 610 in $R_c$ band, 603 in $I_c$ band, respectively.

During the W-D processing, the effective temperature of star 1 is chosen as $T_1=7000$K according to the spectral type determined \citep{Cox2000}. Convective outer envelopes for both components are assumed. The bolometric albedo $A_1=A_2=0.5$ \citep{1969AcA....19..245R} and the values of the gravity-darkening coefficients $g_1=g_2=0.32$ \citep{1967ZA.....65...89L} are used. To account for the limb darkening in detail, logarithmic functions are used. The corresponding bolometric and passband-specific limb-darkening coefficients are chosen from \citet{1993AJ....106.2096V}'s table.
During the calculating, it is found that the solution converges at mode 3, and the adjustable parameters are: the orbital inclination $i$; the mean surface temperature of star 2 ($T_{2}$); the monochromatic luminosity of star 1 ($L_{1B}$, $L_{1V}$, $L_{1R}$ and $L_{1I}$); the dimensionless potential of star 1 ($\Omega_{1}=\Omega_{2}$ in mode 3 for overcontact configuration) and the third light $l_3$. The final photometric solutions are listed in Table 4 and the theoretical light curves are displayed in Fig. 3. The theoretical light curves which haven't been contaminated by the third light is also plotted in Fig. 3 with using the dash lines. The contact configuration of V776 Cas is displayed in Fig. 4.

\begin{table}[!h]
\begin{center}
\caption{Photometric solutions of V776 Cas}\label{phsolutions}
\scriptsize
\begin{tabular}{ccccccccccc}
\hline\hline
Parameters                            & without $l_3$                &  with $l_3$                  & without $l_3$                &  with $l_3$ \\\hline
$T_{1}(K)   $                         & 7000(fixed)                  &  7000(fixed)                 & 6890(fixed)                  &  6890(fixed)\\
$g_{1}$                               & 0.32(fixed)                  &  0.32(fixed)                 & 0.32(fixed)                  &  0.32(fixed) \\
$g_{2}$                               & 0.32(fixed)                  &  0.32(fixed)                 & 0.32(fixed)                  &  0.32(fixed) \\
$A_{1}$                               & 0.50(fixed)                  &  0.50(fixed)                 & 0.50(fixed)                  &  0.50(fixed)  \\
$A_{2}$                               & 0.50(fixed)                  &  0.50(fixed)                 & 0.50(fixed)                  &  0.50(fixed) \\
q ($M_2/M_1$ )                        & 0.130(fixed)                 &  0.130(fixed)                & 0.130(fixed)                 &  0.130(fixed)\\
$i(^{\circ})$                         & 54.2($\pm0.1$)               &  55.4($\pm0.1$)              & 54.3($\pm0.1$)               &  55.4($\pm0.1$)  \\
$\Omega_{in}$                         & 2.0476                       &  2.0476                      & 2.0476                       &  2.0476\\
$\Omega_{out}$                        & 1.9633                       &  1.9633                      & 1.9633                       &  1.9633 \\
$\Omega_{1}=\Omega_{2}$               & 2.0124($\pm0.0009$)          &  1.9932($\pm0.0007$)         & 2.0128($\pm0.0009$)          &  1.9932($\pm0.0007$)\\
$T_{2}(K)$                            & 6865($\pm10$)                &  6920($\pm2$)                & 6749($\pm11$)                &  6796($\pm2$)\\
$\Delta T(K)$                         & 135                          &  80                          & 141                          &  94  \\
$T_{2}/T_{1}$                         & 0.9807($\pm0.0014$)          &  0.9886($\pm0.0003$)         & 0.9795($\pm0.0016$)          &  0.9863($\pm0.0003$)  \\
$L_{1}/L_{T}$ ($B$)                   & 0.8671($\pm0.0002$)          &  0.8573($\pm0.0019$)         & 0.8682($\pm0.0002$)          &  0.8587($\pm0.0019$) \\
$L_{1}/L_{T}$ ($V$)                   & 0.8647($\pm0.0002$)          &  0.8560($\pm0.0018$)         & 0.8656($\pm0.0002$)          &  0.8570($\pm0.0018$) \\
$L_{1}/L_{T}$ ($R_c$)                 & 0.8633($\pm0.0002$)          &  0.8552($\pm0.0020$)         & 0.8640($\pm0.0002$)          &  0.8560($\pm0.0020$) \\
$L_{1}/L_{T}$ ($I_c$)                 & 0.8620($\pm0.0002$)          &  0.8544($\pm0.0028$)         & 0.8626($\pm0.0002$)          &  0.8551($\pm0.0028$)\\
$L_{3}/L'_{T}$ ($B$)                  &                              &  0.1423($\pm0.0099$)         &                              &  0.1452($\pm0.0100$)  \\
$L_{3}/L'_{T}$ ($V$)                  &                              &  0.1338($\pm0.0099$)         &                              &  0.1358($\pm0.0100$) \\
$L_{3}/L'_{T}$ ($R_c$)                &                              &  0.1531($\pm0.0109$)         &                              &  0.1553($\pm0.0110$) \\
$L_{3}/L'_{T}$ ($I_c$)                &                              &  0.1675($\pm0.0152$)         &                              &  0.1699($\pm0.0152$) \\
$r_{1}(pole)$                         & 0.5270($\pm0.0003$)          &  0.5324($\pm0.0007$)         & 0.5269($\pm0.0003$)          &  0.5325($\pm0.0007$)  \\
$r_{1}(side)$                         & 0.5861($\pm0.0004$)          &  0.5948($\pm0.0011$)         & 0.5860($\pm0.0004$)          &  0.5949($\pm0.0011$)\\
$r_{1}(back)$                         & 0.6086($\pm0.0005$)          &  0.6189($\pm0.0013$)         & 0.6084($\pm0.0005$)          &  0.6191($\pm0.0013$) \\
$r_{2}(pole)$                         & 0.2167($\pm0.0003$)          &  0.2233($\pm0.0009$)         & 0.2165($\pm0.0003$)          &  0.2235($\pm0.0009$) \\
$r_{2}(side)$                         & 0.2269($\pm0.0004$)          &  0.2349($\pm0.0011$)         & 0.2269($\pm0.0004$)          &  0.2352($\pm0.0011$) \\
$r_{2}(back)$                         & 0.2725($\pm0.0009$)          &  0.2935($\pm0.0031$)         & 0.2721($\pm0.0009$)          &  0.2944($\pm0.0032$) \\
$f$                                   & $41.8\,\%$($\pm$1.1\,\%$$)   &  $64.6\,\%$($\pm$2.9\,\%$$)  & $41.4\,\%$($\pm$1.1\,\%$$)   &  $65.4\,\%$($\pm$2.9\,\%$$) \\
$\Sigma{\omega(O-C)^2}$               & 0.049755                     &  0.042135                    & 0.049924                     &  0.042369 \\
\hline
\end{tabular}
\end{center}
{\footnotesize Notes:} \footnotesize $L_{T}=L_{1}+L_{2}, L'_{T}=L_{1}+L_{2}+L_{3}$
\end{table}

\begin{figure}[!h]
\begin{center}
\includegraphics[width=14cm]{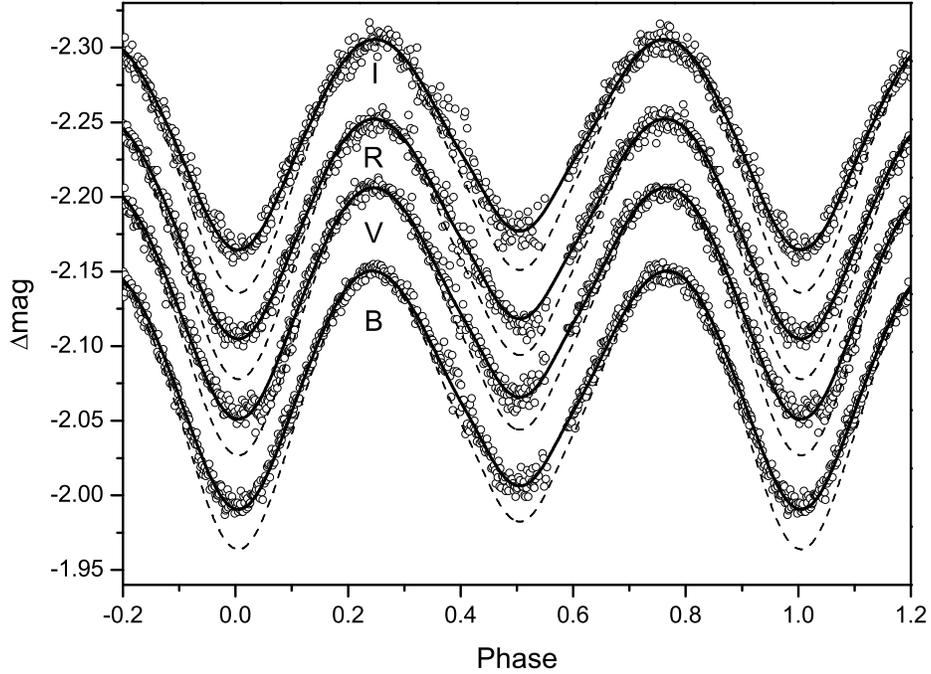}
\caption{Observed (open circles) and theoretical (solid lines) light curves in the $B$ $V$ $R_c$ and $I_c$ bands for V776 Cas. The standard deviation of the fitting residuals are 0.005 mag for $B$ band, 0.006 mag for $V$ band, 0.006 mag for $R_c$ band and 0.006 mag for $I_c$ band, respectively. The dash lines represent the theoretical light curves without third light.}
\end{center}
\end{figure}

\begin{figure}[!h]
\begin{center}
\includegraphics[width=14cm]{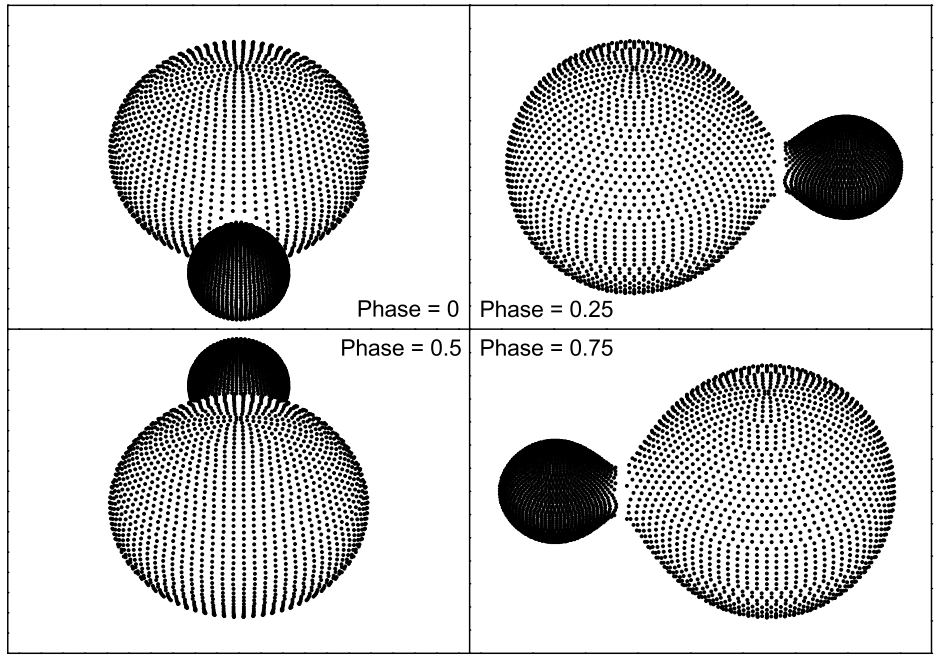}
\caption{Contact configurations of V776 Cas at phase 0.0, 0.25, 0.5, 0.75,}
\end{center}
\end{figure}

It has to be mentioned that the mass ratio q ($M_2/M_1$ ) and the effective temperature of the primary star ($T_1$) are very important input parameters when the W-D calculations are carried out. The newly obtained mass ratio of V776 Cas was $q = 0.1306\pm0.0046$ with a very high accuracy \citep{2012JKAS...45....1G} , which was consistent with that of \citet{2001AJ....122.1974R}, so we chose $q = 0.130$ in our calculations. And another solution of $T_1=6890$K are also obtained and listed in Table 4. We can conclude that it makes almost no difference to the results when the the effective temperature are set as $T_1 = 7000$K and $T_1 = 6890$K. Therefore, we use the solution of $T_1 = 7000$K as our final results hereafter in the paper.

\section{Discussions and Conclusions}
The light curve solutions indicate that V776 Cas is an A-subtype overcontact binary system with a very high contact degree ($ f=64.6\,\%$) and an extremely low mass ratio ($q=0.130$), which indicate that it is at the final evolutionary stage of cool short-period binary. It may merge into a single rapid-rotation star, which may be the progenitor of blue straggler or FK Com-type star \citep{2015AJ....150...83Z}. The two components have nearly the same surface temperature ($\Delta T = 80K$) in spite of their quite different mass and radii. It suggests that the system is under thermal contact. Considering the orbital inclination ($i = 55.4^{\circ}$) of ours and the mass function given by \citet{2001AJ....122.1974R}: $(M_1+M_2)sin^3i = 0.975\pm0.026M_\odot$, we can easily calculate the mass of the two components to be $M_1 = 1.55(\pm0.04)M_\odot$, $M_2 = 0.20(\pm0.01)M_\odot$. The absolute physical parameters of the two components in V776 Cas are listed in Table 5.

As the spectroscopic search carried out by \citet{2006AJ....132..650D} confirmed that V776 Cas was a triple system with a solar-type tertiary component ($T_3 = 6100K$), the third light ($l_3$) is also included as an adjustable parameter during the photometric processing. The results suggest that the third light contributes nearly $15\,\%$ of the total luminosity and it reduces the depths of the primary and secondary minimum apparently as shown in Fig. 3. The third component mainly radiates in $R_c$ and $I_c$ bands. According to the third light values in $R_c$ and $I_c$ filters listed in Table 4, the color index of the tertiary component are calculated to be $R_c - I_c = 0.32$, which corresponds to a spectral type of G0V. It is consistent with the result of spectroscopic fitting. The mass of the third component is estimated to be $M_3 = 1.04(\pm0.03)M_\odot$ according to the mass ratio ($M_3/M_1 = 0.67$) obtained by \citet{2006AJ....132..650D}. By assuming a circular orbit ($e=0.0$), the projected radius of the orbit that the eclipsing binary rotates around the barycenter of the triple system is calculated with the following equation,
\begin{equation}
\begin{array}{lll}
a'_{12}\sin i'=A_3 \times c,
\end{array}
\end{equation}
where $A_3$ is the amplitude of the $O-C$ oscillation and $c$ is the speed of light, i.e. , $a'_{12}\sin i'=1.88(\pm0.02)AU$. The mass function and the mass of the tertiary companion are computed with the following equation,
\begin{equation}
\begin{array}{lll}
f(m)=\frac{4\pi^2}{GP^2_3}\times(a'_{12}\sin i')^3=\frac{(M_3\sin i')^3}{(M_1+M_2+M_3)^2},
\end{array}
\end{equation}
where $G$ and $P_3$ are the gravitational constant and the period of the $(O-C)_2$ oscillation. The orbital inclination of the third component is calculated to be $i' = 33.1^{\circ}$. The distance of the binary system to the barycenter of the triple system is calculated to be $a'_{12} = 3.45AU$.

Over the past decade, light curve solutions of V776 Cas have been achieved by \citet{2004NewA....9..425D} (\citet{2004ASPC..318..192E}), \citet{2005AcA....55..389Z} and \citet{2007ASPC..362...82O}. Their results are listed in Table 6. Our results are consistent with those of \citet{2004NewA....9..425D}'s. The solutions of \citet{2005AcA....55..389Z} and \citet{2007ASPC..362...82O} gave consistent results. Compared with our results, they obtained a lower orbital inclination and a higher contact degree. The effective temperature of the two components were very close to each other and the third light ($l_3$) was almost negligible. These may be caused by the different mass ratio used in W-D calculation or due to the effect of different third light contamination. However, magnetic activities and mass transfer between the two components may also response to the difference among the solutions since there is a late-type component in the overcontact binary system and \citet{2007ASPC..362...82O} has used a spot model to modeling the light curves. As both spectroscopic observation and $O - C$ curve analysis confirm the existence of a solar-type close-in tertiary in the close binary system, our results with a contribution of third light may be much more acceptable.

The formation and evolutionary scenario of W UMa-type binaries are still an unsolved problem in astrophysics. According to the method established by \citet{2013MNRAS.430.2029Y}, the initial mass of the two components in V776 Cas are calculated to be $M_{2i} = 2.13(\pm0.04)M_\odot$ and $M_{1i} = 0.86(\pm0.10)M_\odot$. The initial mass of the present less massive one ($M_{2i}$) is within the mass range pertaining to the A-subtype close binaries. The initial mass of the present primary one ($M_{1i}$) indicates that it is a late-type star, which has played a very important role for early angular momentum evolution of close binaries as magnetic braking is supposed to be the main mechanism for such stars with convective envelope. For V776 Cas, the mass loss of the secondary star is calculated to be $1.93M_\odot$ and the mass gained by the primary component is $0.69M_\odot$, which also confirm the conclusion that only one-third of the mass lost by the secondary component is transferred to the primary component \citep{2013MNRAS.430.2029Y}.

V776 Cas is the brighter member of the visual binary ADS 1485. The close binary system is even confirmed to be a triple system with a solar-type third component orbiting around the close binary system. Thus, it is actually a quadruple system. \citet{2006A&A...450..681T} surveyed a sample of 165 solar-type spectroscopic binaries and found that the fraction of spectroscopic binaries with additional companions is $63\,\% \pm ¡À 5\,\%$. The fraction had strong correlation with the periods ($P$) of binary systems, which reached  $96\,\%$ for $P < 3$d and dropped to $34\,\%$ for
$P > 12$d. \citet{2013ApJS..209...13Q,2014AJ....148...79Q} thought that the existence of an additional stellar component in the binary system may play an important role for the formation and evolution by removing angular momentum from the central binary system during the early dynamical interaction or late evolution. It is possible that third-body interactions in the birth environment may help to accelerate the orbital evolution of the central binary system. Angular momentum is drained from the inner close pair either by the ejection of the tertiary companion \citep{2004A&A...414..633G} or through the Kozai mechanism \citep{1962AJ.....67..591K,2007ApJ...669.1298F}. The angular momentum and orbital period of the binary system will decrease, and the initially detached binary system evolves into the contact configuration during their main sequence evolutionary stage. V776 Cas is an important target for testing theories of W UMa-type bianries' formation and stellar dynamical interaction and evolution.

\begin{table}[!h]
\caption{Absolute parameters of the two components in V776 Cas}\label{absolute}
\begin{center}
\small
\begin{tabular}{lllllllll}
\hline
Parameters                        &Primary                         & Secondary          \\
\hline
$M$                               & $1.55(\pm0.04)M_\odot$         & $0.20(\pm0.01)M_\odot$         \\
$R$                               & $1.71(\pm0.15)R_\odot$         & $0.73(\pm0.06)R_\odot$         \\
$L$                               & $6.93(\pm0.72)L_\odot$         & $1.17(\pm0.12)L_\odot$         \\
\hline
\end{tabular}
\end{center}
\end{table}

\begin{table}[!h]
\begin{center}
\caption{Photometric solutions of V776 Cas for the past decade}\label{phsolutions}
\scriptsize
\begin{tabular}{cccccccccccc}
\hline\hline
Parameters                           & \citet{2004NewA....9..425D} & \citet{2005AcA....55..389Z} & \citet{2007ASPC..362...82O}& The present work \\\hline
$T_{1}(K)   $                        & 6890                        & 6700                        & 7047                       & 7000\\
q ($M_2/M_1$ )                       & 0.130                       & 0.138                       & 0.145($\pm0.001$)          & 0.130\\
$i(^{\circ})$                        & 55.8($\pm0.2$)              & 52.5($\pm0.9$)              & 53.584($\pm0.134$)         & 55.4($\pm0.1$)\\
$\Omega_{1}=\Omega_{2}$              & 1.9984                      & 2.001($\pm0.008$)           & 2.0127($\pm0.0019$)        & 1.9932($\pm0.0007$)\\
$f$                                  & $58.39\,\%$                 & $77\,\%$                    & $81.9\,\%$                 & $64.6\,\%$($\pm$2.9\,\%$$) \\
$T_{2}(K)$                           & 6620($\pm46$)               & 6725($\pm90$)               & 7004($\pm39$)              & 6920($\pm2$)\\
$\Delta T(K)$                        & 270                         & 25                          & 43                         & 80 \\
$T_{2}/T_{1}$                        & 0.9608($\pm0.0067$)         & 1.0037($\pm0.0134$)         & 0.9939($\pm0.0055$)        & 0.9886($\pm0.0003$)\\
$L_{1}/L_{T}$ ($U$)                  &                             & 0.8386($\pm0.0053$)         &                            &                    \\
$L_{1}/L_{T}$ ($B$)                  &                             & 0.8380($\pm0.0059$)         & 0.8362                     & 0.8573($\pm0.0019$)\\
$L_{1}/L_{T}$ ($V$)                  & 0.866                       & 0.8390($\pm0.0056$)         & 0.8357                     & 0.8560($\pm0.0018$)\\
$L_{1}/L_{T}$ ($R_c$)                &                             & 0.8392($\pm0.0052$)         & 0.8354                     & 0.8552($\pm0.0020$) \\
$L_{1}/L_{T}$ ($I_c$)                &                             &                             &                            & 0.8544($\pm0.0028$)\\
$L_{3}/L'_{T}$ ($U$)                 &                             & 0.0071($\pm0.0031$)         &                            &                     \\
$L_{3}/L'_{T}$ ($B$)                 &                             & 0.0074($\pm0.0032$)         & 0.0174                     & 0.1423($\pm0.0099$) \\
$L_{3}/L'_{T}$ ($V$)                 & 0.136                       & 0.0061($\pm0.0033$)         & 0.0162                     & 0.1338($\pm0.0099$)\\
$L_{3}/L'_{T}$ ($R_c$)               &                             & 0.0041($\pm0.0031$)         & 0.0152                     & 0.1531($\pm0.0109$)\\
$L_{3}/L'_{T}$ ($I_c$)               &                             &                             &                            & 0.1675($\pm0.0152$)\\
$\theta(^{\circ})$                   &                             &                             & 90                         &                 \\
$\psi(^{\circ})$                     &                             &                             & 274.90($\pm0.62$)          &                 \\
$r$(rad)                             &                             &                             & 10                         &                    \\
$T_f$                                &                             &                             & 0.980($\pm0.010$)          &                  \\
$M_{1}(M_\odot)$                     & 1.63                        & 1.750($\pm0.040$)           & 1.71                       & 1.55($\pm0.04$)\\
$M_{2}(M_\odot)$                     & 0.21                        & 0.242($\pm0.017$)           & 0.25                       & 0.20($\pm0.01$)\\
$R_{1}(R_\odot)$                     & 1.73                        & 1.821($\pm0.017$)           & 1.77                       & 1.71($\pm0.15$)\\
$R_{2}(R_\odot)$                     & 0.74                        & 0.748($\pm0.012$)           & 0.81                       & 0.73($\pm0.06$)\\
$L_{1}(L_\odot)$                     &                             & 5.90($\pm0.11$)             & 6.83                       & 6.93($\pm0.72$)\\
$L_{2}(L_\odot)$                     &                             & 1.01($\pm0.06$)             & 1.39                       & 1.17($\pm0.12$)\\
$a_{orb}(R_\odot)$                   & 2.985                       & 3.07($\pm0.03$)             & 3.05                       & 2.94($\pm0.03$)\\
\hline
\end{tabular}
\end{center}
{\footnotesize Notes:} \footnotesize $L_{T}=L_{1}+L_{2}, L'_{T}=L_{1}+L_{2}+L_{3}$
\end{table}

\acknowledgments{ We thank the anonymous referee for useful comments and suggestions that have improved the quality of the manuscript. This work is supported by the Chinese Natural Science Foundation (Grant No. 11133007 and 11325315), the Strategic Priority Research Program ``The Emergence of Cosmological Structure'' of the Chinese Academy of Sciences (Grant No. XDB09010202) and the Science Foundation of Yunnan Province (Grant No. 2012HC011). New CCD photometric observations of V776 Cas were obtained with the 60cm and the 1.0m telescopes at the Yunnan Observatories, and the 50cm and 85cm telescopes in Xinglong Observation base in China.}


\begin{thebibliography}{}
\bibitem[Basturk et al.(2014)]{2014IBVS.6125....1B} Basturk, O., Bahar, E., Senavci, H.~V., et al.\ 2014, Information Bulletin on Variable Stars, 6125, 1
\bibitem[Cox(2000)]{Cox2000} Cox, A. N. 2000, Allen$^\prime$s Astrophysical Quantities (4th ed.; NewYork: Springer)
\bibitem[Csizmadia et al.(2006)]{2006IBVS.5736....1C} Csizmadia, S., Klagyivik, P., Borkovits, T., et al.\ 2006, Information Bulletin on Variable Stars, 5736, 1
\bibitem[D'Angelo et al.(2006)]{2006AJ....132..650D} D'Angelo, C., van Kerkwijk, M.~H., \& Rucinski, S.~M.\ 2006, \aj, 132, 650
\bibitem[Djura{\v s}evi{\'c} et al.(2004)]{2004NewA....9..425D} Djura{\v s}evi{\'c}, G., Albayrak, B., Selam, S.~O., Erkapi{\'c}, S., \& {\c S}enavc{\i}, H.~V.\ 2004, \na, 9, 425
\bibitem[Duerbeck(1997)]{1997IBVS.4513....1D} Duerbeck, H.~W.\ 1997, Information Bulletin on Variable Stars, 4513, 1
\bibitem[Elmasli et al.(2004)]{2004ASPC..318..192E} Elmasli, A., Tanriverdi, T., Albayrak, B., Selam, S.~O., \& Djurasevic, G.\ 2004, Spectroscopically and Spatially Resolving the Components of the Close Binary Stars, 318, 192
\bibitem[Fabrycky \& Tremaine(2007)]{2007ApJ...669.1298F} Fabrycky, D., \& Tremaine, S.\ 2007, \apj, 669, 1298
\bibitem[Flannery(1976)]{1976ApJ...205..217F} Flannery, B.~P.\ 1976, \apj, 205, 217v
\bibitem[Ghaderi et al.(2012)]{2012JKAS...45....1G} Ghaderi, K., Pirkhedri, A., Rostami, T., Khodamoradi, S., \& Fatahi, H.\ 2012, Journal of Korean Astronomical Society, 45, 1
\bibitem[Goodwin et al.(2004)]{2004A&A...414..633G} Goodwin, S.~P., Whitworth, A.~P., \& Ward-Thompson, D.\ 2004, \aap, 414, 633
\bibitem[Gomez-Forrellad et al.(1999)]{1999IBVS.4702....1G} Gomez-Forrellad, J.~M., Garcia-Melendo, E., Guarro-Flo, J., Nomen-Torres, J., \& Vidal-Sainz, J.\ 1999, Information Bulletin on Variable Stars, 4702, 1
\bibitem[Hilditch et al.(1988)]{1988MNRAS.231..341H} Hilditch, R.~W., King, D.~J., \& McFarlane, T.~M.\ 1988, \mnras, 231, 341
\bibitem[Ho{\v n}kov{\'a} et al.(2013)]{2013OEJV..160....1H} Ho{\v n}kov{\'a}, K., Jury{\v s}ek, J., Lehk{\'y}, M., et al.\ 2013, Open European Journal on Variable Stars, 160, 1
\bibitem[Hubscher et al.(2010)]{2010IBVS.5941....1H} Hubscher, J., Lehmann, P.~B., Monninger, G., Steinbach, H.-M., \& Walter, F.\ 2010, Information Bulletin on Variable Stars, 5941, 1
\bibitem[Hubscher \& Lehmann(2012)]{2012IBVS.6026....1H} Hubscher, J., \& Lehmann, P.~B.\ 2012, Information Bulletin on Variable Stars, 6026, 1
\bibitem[Hubscher \& Lehmann(2013)]{2013IBVS.6070....1H} Hubscher, J., \& Lehmann, P.~B.\ 2013, Information Bulletin on Variable Stars, 6070, 1
\bibitem[Kozai(1962)]{1962AJ.....67..591K} Kozai, Y.\ 1962, \aj, 67, 591
\bibitem[Krajci(2005)]{2005IBVS.5592....1K} Krajci, T.\ 2005, Information Bulletin on Variable Stars, 5592, 1
\bibitem[Liao \& Qian(2010)]{2010MNRAS.405.1930L} Liao, W.-P., \& Qian, S.-B.\ 2010, \mnras, 405, 1930
\bibitem[Lindegren(1997)]{1997ESASP.402...13L} Lindegren, L.\ 1997, Hipparcos - Venice '97, 402, 13
\bibitem[Lucy(1967)]{1967ZA.....65...89L} Lucy, L.~B.\ 1967, \zap, 65, 89
\bibitem[Lucy(1976)]{1976ApJ...205..208L} Lucy, L.~B.\ 1976, \apj, 205, 208
\bibitem[Mestel(1968)]{1968MNRAS.138..359M} Mestel, L.\ 1968, \mnras, 138, 359
\bibitem[Nelson(2005)]{2005IBVS.5602....1N} Nelson, R.~H.\ 2005, Information Bulletin on Variable Stars, 5602, 1
\bibitem[Nelson(2007)]{2007IBVS.5760....1N} Nelson, R.~H.\ 2007, Information Bulletin on Variable Stars, 5760, 1
\bibitem[Nelson(2009)]{2009IBVS.5875....1N} Nelson, R.~H.\ 2009, Information Bulletin on Variable Stars, 5875, 1
\bibitem[Nelson(2011)]{2011IBVS.5966....1N} Nelson, R.~H.\ 2011, Information Bulletin on Variable Stars, 5966, 1
\bibitem[Okamoto \& Sato(1970)]{1970PASJ...22..317O} Okamoto, I., \& Sato, K.\ 1970, \pasj, 22, 317
\bibitem[Oh et al.(2007)]{2007ASPC..362...82O} Oh, K.-D., Kim, C.-H., Kim, H.-I., \& Lee, W.-B.\ 2007, The Seventh Pacific Rim Conference on Stellar Astrophysics, 362, 82
\bibitem[Parimucha et al.(2007)]{2007IBVS.5777....1P} Parimucha, S., Vanko, M., Pribulla, T., et al.\ 2007, Information Bulletin on Variable Stars, 5777, 1
\bibitem[Parimucha et al.(2009)]{2009IBVS.5898....1P} Parimucha, S., Dubovsky, P., Baludansky, D., et al.\ 2009, Information Bulletin on Variable Stars, 5898, 1
\bibitem[Parimucha et al.(2011)]{2011IBVS.5980....1P} Parimucha, S., Dubovsky, P., Vanko, M., et al.\ 2011, Information Bulletin on Variable Stars, 5980, 1
\bibitem[Parimucha et al.(2013)]{2013IBVS.6044....1P} Parimucha, S., Dubovsky, P., \& Vanko, M.\ 2013, Information Bulletin on Variable Stars, 6044, 1
\bibitem[Qian(2001)]{2001MNRAS.328..914Q} Qian, S.\ 2001, \mnras, 328, 914
\bibitem[Qian et al.(2013)]{2013ApJS..209...13Q} Qian, S.-B., Liu, N.-P., Li, K., et al.\ 2013, \apjs, 209, 13
\bibitem[Qian et al.(2014)]{2014AJ....148...79Q} Qian, S.-B., Zhou, X., Zola, S., et al.\ 2014, \aj, 148, 79
\bibitem[Robertson \& Eggleton(1977)]{1977MNRAS.179..359R} Robertson, J.~A., \& Eggleton, P.~P.\ 1977, \mnras, 179, 359
\bibitem[Rucinski et al.(2001)]{2001AJ....122.1974R} Rucinski, S.~M., Lu, W., Mochnacki, S.~W., Og{\l}oza, W., \& Stachowski, G.\ 2001, \aj, 122, 1974
\bibitem[Ruci{\'n}ski(1969)]{1969AcA....19..245R} Ruci{\'n}ski, S.~M.\ 1969, \actaa, 19, 245
\bibitem[Schatzman(1962)]{1962AnAp...25...18S} Schatzman, E.\ 1962, Ann. Astrophysique, 25, 18
\bibitem[Tanriverdi et al.(2003)]{2003IBVS.5407....1T} Tanriverdi, T., Kutdemir, E., Elmasli, A., et al.\ 2003, Information Bulletin on Variable Stars, 5407, 1
\bibitem[Tokovinin et al.(2006)]{2006A&A...450..681T} Tokovinin, A., Thomas, S., Sterzik, M., \& Udry, S.\ 2006, \aap, 450, 681
\bibitem[Tutukov et al.(2004)]{2004ARep...48..219T} Tutukov, A.~V., Dremova, G.~N., \& Svechnikov, M.~A.\ 2004, Astronomy Reports, 48, 219
\bibitem[vant Veer(1979)]{1979A&A....80..287V} vant Veer, F.\ 1979, \aap, 80, 287
\bibitem[Van Hamme(1993)]{1993AJ....106.2096V} Van Hamme, W.\ 1993, \aj, 106, 2096
\bibitem[Van Hamme \& Wilson(2007)]{Van2007} Van Hamme, W., \& Wilson, R.~E.\ 2007, \apj, 661, 1129
\bibitem[Vilhu(1982)]{1982A&A...109...17V} Vilhu, O.\ 1982, \aap, 109, 17
\bibitem[Webbink(1976)]{1976ApJS...32..583W} Webbink, R.~F.\ 1976, \apjs, 32, 583
\bibitem[Wilson(1979)]{Wilson1979} Wilson, R. E. 1979, ApJ, 234, 1054
\bibitem[Wilson(1990)]{Wilson1990} Wilson, R. E. 1990, ApJ, 356, 613
\bibitem[Wilson(2008)]{Wilson2008} Wilson, R. E.,  2008, ApJ, 672, 575
\bibitem[Wilson(2012)]{Wilson2012} Wilson, R. E., 2012, AJ, 144, 73
\bibitem[Wilson et al.(1971)]{Wilson1971} Wilson, R. E., Devinney E.J. 1971, ApJ, 166, 605
\bibitem[Wilson et al.(2010)]{Wilson2010} Wilson, R. E., Van, Hamme. W., Terrell, D., 2010, ApJ, 723, 1469
\bibitem[Y{\i}ld{\i}z \& Do{\u g}an(2013)]{2013MNRAS.430.2029Y} Y{\i}ld{\i}z, M., \& Do{\u g}an, T.\ 2013, \mnras, 430, 2029
\bibitem[Y{\i}ld{\i}z(2014)]{2014MNRAS.437..185Y} Y{\i}ld{\i}z, M.\ 2014, \mnras, 437, 185
\bibitem[Yilmaz et al.(2009)]{2009IBVS.5887....1Y} Yilmaz, M., Basturk, O., Alan, N., et al.\ 2009, Information Bulletin on Variable Stars, 5887, 1
\bibitem[Zasche et al.(2014)]{2014IBVS.6114....1Z} Zasche, P., Uhlar, R., Kucakova, H., Svoboda, P., \& Masek, M.\ 2014, Information Bulletin on Variable Stars, 6114, 1
\bibitem[Zhou et al.(2015)]{2015AJ....150...83Z} Zhou, X., Qian, S.-B., Liao, W.-P., et al.\ 2015, \aj, 150, 83
\bibitem[Zola et al.(2005)]{2005AcA....55..389Z} Zola, S., Kreiner, J.~M., Zakrzewski, B., et al.\ 2005, \actaa, 55, 389
\end{thebibliography}
\end{document}